\documentclass[reqno]{amsart}
\usepackage{amsmath}

\def\beq{\begin{equation}}
\def\eeq{\end{equation}}

\def\half{{\textstyle{1\over2}}}
\def\two3{{\textstyle{2\over3}}}

\def\beq{\begin{equation}}
\def\eeq{\begin{equation}}

\newcommand{\dd}{\mathrm{d}}
\newcommand{\pp}{\ .}
\newcommand{\Tr}{\mathop{\mathrm{Tr}}}
\newcommand{\bbox}[1]{\mathbf{#1}}
\let\cal\mathcal

\textwidth=\hsize

\begin{document}

\hoffset=-0.5truein
%


\title{%
Dimensionally Reduced Chern-Simons Terms and their Solitons}

\author{
R.~Jackiw}\address{
RJ: Center for Theoretical Physics\\
 Massachusetts Institute of
Technology\\
 Cambridge, MA ~02139--4307}
\thanks{Soliton Symposium, Durham, England, July 1998.}
\author{So-Young Pi}\address{S-\negthinspace YP:
Physics Department\\ Boston University\\ Boston, MA ~02215}
\thanks{\leavevmode\llap{``}Quantum Field Theory Perspective and
Prospective,'' Les Houches, France, June 1998.\\[1ex] Research supported in part by
funds provided by  the U.S.~Department of Energy (D.O.E.) under contracts
\#DE-FC02-94ER40818 and \#DE-FG02-91ER40676.\\[1ex]  MIT-CTP-2771\qquad
BU-HEP-98-23\qquad hep-th/9808036
\hfill August 1998}

\begin{abstract}%
\noindent%
We consider models in which nonrelativistic matter fields interact with gauge fields
whose dynamics are governed by the Chern-Simons term.  The relevant equations of
motion are derived and reduced dimensionally in time or in space.  Interesting solitonic
equations emerge and their solutions are described.  Finally, we consider a
Chern-Simons term in three-dimensional Euclidean space, reduced by spherical symmetry,
and we discuss its effect on monopole and instanton solutions.
\end{abstract}

\maketitle

\section{Introduction}

In (2+1)-dimensions, the possibility of describing gauge theories with a Chern-Simons
(C-S) kinetic term, rather than with a Yang-Mills kinetic term, leads to physically and
mathematically interesting consequences.  The C-S term is given by
\begin{equation}
W(A) = -\frac{1}{16 \pi^2} \int \dd^3 x\, \epsilon^{\alpha \beta\gamma} \Tr 
(A_\alpha \partial_\beta A_\gamma + \two3 A_\alpha A_\beta A_\gamma)
\label{eq:1jp}
\end{equation}
where $\epsilon^{\alpha \beta\gamma}$ is the totally antisymmetric tensor and the
gauge field $A_\mu = A_\mu^a T^a$ takes values in a finite-dimensional representation of
a Lie algebra, with generators $T^a$ satifying $[T^a, T^b] = f^{ab} {}_cT^c$.  In the
Abelian theory the trilinear term vanishes since $A_\mu$'s commute.

$W(A)$ possesses the important property of being invariant against infinitesimal gauge
transformations, while changing under large gauge transformations by the integer
winding number $n$ of the group element $g$ that effects the transformation:
\begin{subequations}
\begin{eqnarray}
A_\mu &\to& A_\mu^g \equiv g^{-1} A_\mu g + g^{-1} \partial_\mu g 
\label{eq:2ajp} \\
W(A) &\to& W(A^g) = W(A) + n \pp
\label{eq:2bjp}
\end{eqnarray}
(We have assumed that no surface terms involving $A$ contribute.)
\begin{equation}
n=-\frac{1}{48 \pi^2} \int  \dd^3 x\, \epsilon^{\alpha \beta\gamma} \Tr 
(g^{-1} \partial_\alpha gg^{-1} \partial_\beta gg^{-1} \partial_\gamma g)
\label{eq:2cjp}
\end{equation}
\end{subequations}%
Therefore, when the C-S term $W(A)$ is used in the action for (2+1)-dimensional
gauge fields of a non-Abelian group, the coefficient of $W(A)$ must be properly
quantized in order to ensure gauge invariance of the quantum theory.  Then the shift in
$W(A)$ under a gauge transformation is not seen in the phase exponential of the
action.\cite{ref:1,ref:2}

Another important feature of the C-S term is that it describes a topological quantity in
the sense that there is no explicit dependence on the space-time metric.  Thus when
$W(A)$ is used in the action it does not contribute to the energy-momentum tensor,
which is obtained by varying the action with respect to the metric tensor.

When the C-S term is coupled to matter fields in a gauge covariant manner we have an
action of the following form:
\begin{equation}
I=8\pi^2 \kappa W(A) - \int \dd^3 x\,  A_\mu J^\mu +I_{\rm matter} \pp
\label{eq:3jp}
\end{equation}
The matter action may describe matter fields with relativistic or nonrelativistic
dynamics.  The total action is then either Lorentz or Galilean invariant since the C-S
term is topological, that is,  invariant against \emph{all} coordinate tranformations. 
Because the variation of $W(A)$ with respect to $A_\mu$ is $\frac{1}{16 \pi^2}
\epsilon^{\mu
\alpha\beta} F_{\alpha\beta}$, the action in Eq.~(\ref{eq:3jp}) produces a field-current
indentity
\begin{equation}
\frac{\kappa}{2} \epsilon^{\mu \alpha\beta}F_{\alpha\beta} = J^\mu 
\label{eq:4jp}
\end{equation}
which tells us that gauge fields are not dynamical, but they are completely determined
by matter variables.  Therefore, the resulting matter equations are self-contained and
highly nonlinear.

When matter field dynamics are nonrelativistic, we have an interesting class of C-S
gauge models described by the action
\begin{equation}
I=8\pi^2\kappa W(A) + \int \dd^3 x\,  \Bigl\{ i \psi^\dagger D_t \psi -
\frac{1}{2m} (\bbox{D} \psi)^\dagger (\bbox{D}\psi) -V (\rho) \Bigr\} \pp
\label{eq:5jp}
\end{equation}
$\psi$ is a scalar multiplet in some definite representation of a non-Abelian gauge
group and $(D_t, \bbox{D})$ are (temporal, spatial) gauge covariant derivatives. 
$V(\rho)$ is a scalar potential describing matter self-interaction with $\rho^a = -i
\psi^\dagger T^a
\psi$, where $T^a$ represents anti-Hermitian generators of the gauge algebra in the
representation of $\psi$.

When we choose $V (\rho)$ to be 
\begin{equation}
V(\rho) = -\half g \rho^a \rho_a
\label{eq:6jp}
\end{equation}
the matter field equation that emerges from the action (\ref{eq:5jp}) is a ``gauged''
nonlinear Schr\"odinger equation
\begin{equation}
i \partial_t \psi = -\half \bbox{D}^2 \psi - i A^0 \psi + ig\rho\psi 
\label{eq:7jp}
\end{equation}
where $(A^0, \bbox{A})$ are given by the field-current identity (\ref{eq:4jp}):
\begin{subequations}
\begin{align}
B^a &= - \frac{1}{\kappa} \rho^a \quad {\rm (Gauss'~law)}
\label{eq:8ajp} \\
E^{ia} &= \frac{1}{\kappa}\epsilon^{ij} J_j^a
\label{eq:8bjp}
\end{align}
\end{subequations}%
with the covariantly conserved current density given by
\begin{subequations}
\begin{align}
J_0^a &= \rho^a
\label{eq:9ajp} \\
J_i^a &= \half \big(\psi^\dagger T^a D_i \psi - (D_i \psi)^\dagger T^a \psi\big)
\label{eq:9bjp}
\end{align}
\end{subequations}%
In these lectures we  discuss models belonging to the general class (\ref{eq:5jp}),
with both Abelian and non-Abelian gauge groups.

\section{Dimensional Reduction of Chern-Simons Models}

\subsection{\sl Self-dual theories and reduction to Liouville and Toda equations}

One obvious reduction for the dynamics of (\ref{eq:5jp}) is to reduce in \emph{time}. 
When we choose the scalar potential $V (\rho)$ to be quartic as given in
(\ref{eq:6jp}), with a particular strength $g=\frac{1}{m\kappa}$ and $\kappa>0$, we
obtain a new class of self-dual theories to which one can apply Bogomolny procedures. 
The second-order static equations of (\ref{eq:7jp}) are satisfied by solutions to  first-order self-dual equations, and the Gauss law constraint (\ref{eq:8ajp}). 
\begin{subequations}
\begin{align}
D_i\psi &= i \epsilon_{ij} D_j \psi
\label{eq:10ajp} \\
B &= - \frac{1}{\kappa} \rho \pp
\label{eq:10bjp}
\end{align}
\end{subequations}%
These coupled equations then can be combined into the completely integrable Liouville
equation (Abelian case) or Toda equation $\big($SU(N) case$\big)$, with well-known
soliton solutions.  This subject is an old one, well reviewed in the existing 
literature\cite{ref:3,ref:4}, so we shall not dwell on it.

\subsection{\sl Reduction to nonlinear Schr\"odinger equation}

Now we shall describe in detail a reduction to one \emph{spatial} dimension, which
results in an interesting reformulation of the nonlinear Schr\"odinger equation.  On the
plane, with coordinates
$(x,y)$, we suppress all $y$-dependence and rename $A_y$ as $B$.  Then, in
the Abelian case, the action (\ref{eq:5jp}) becomes
\begin{equation}
I = \int \dd t\, \dd x\, \left\{-\kappa BF + i\psi ^\ast D_t \psi - \frac{1}{2m}
|D\psi |^2  - \frac{1}{2m} B^2\rho - V(\rho)\right\} \pp
\label{eq:11}
\end{equation}
The ``kinetic" gauge field term is the so-called ``$B$-$F$" expression where
$F=\half
\epsilon^{\mu\nu} F_{\mu\nu} = -\dot{a}-A^\prime_0$.  [We have
renamed $A_1$ as $-a$, and dot/dash refer to differentiation with respect to
time/space, that is,  $(t/x)$.  The covariant derivatives read $D_t\psi =
\dot{\psi}+iA_0\psi, D\psi=\psi^\prime-ia\psi$.  Recall that in the Abelian application,
the C-S coefficient is not quantized.]  Evidently the two-dimensional $B$-$F$ quantity is
a dimensional reduction of the C-S expression.

Because (\ref{eq:11}) is first-order in time derivatives, the action is already
in canonical form, and may be analyzed using the symplectic Hamiltonian
procedure.\cite{ref:5}  We present (\ref{eq:11}) as
\begin{align}
I &=  \int\dd t\, \dd x  \left\{\kappa B \dot{a} + i\psi^\ast \dot{\psi} - A_0
\left(\kappa B^\prime + \rho\right) - \frac{1}{2m} \bigl|(\partial_x-ia)\psi\bigr|^2 -
\frac{1}{2m} B^2 \rho - V(\rho)\right\} \nonumber\\
&=  \int\dd t\, \dd x  \left\{\kappa B\dot{a} + i\psi^\ast\dot{\psi}-A_0
\left(\kappa B^\prime+\rho\right) - \frac{1}{2m}
\bigl|(\partial_x-ia\pm B)\psi\bigr|^2 \mp\frac{1}{2m}  B^\prime\rho - V(\rho) \right\}
\label{eq:12}
\end{align}
$A_0$ is a Lagrange multiplier, enforcing the Gauss law, which in this theory requires
\begin{subequations}
\begin{align}
B^\prime &= - \frac{1}{\kappa} \rho 
\label{eq:13a}
\intertext{or equivalently}
B(x) &= -\frac{1}{2\kappa} \int \dd \tilde{x} \, \epsilon(x-\tilde{x})\rho(\tilde{x}) \pp
\label{eq:13b}
\end{align}
\end{subequations}%
[The Green's function, uniquely determined by parity invariance, is the
Heaviside $\pm 1$ step.]  Thus, after we eliminate $B$, (\ref{eq:12}) involves 
a spatially nonlocal Lagrangian.
\begin{subequations}
\begin{align} 
L &= - \half  \int \dd x\, \dd \tilde{x}\,
\dot{a}(x)\epsilon(x-\tilde{x})\rho(\tilde{x})+ \int \dd x\, i \psi^\ast\dot{\psi} 
\nonumber \\  
&\qquad{} -  \frac{1}{2m} \int \dd x\, \biggl| \Bigl(\partial_x - ia \mp
\frac{1}{2\kappa} \int \dd  \tilde{x}\,
\epsilon(x-\tilde{x})\rho(\tilde{x})\Bigr)\psi(x)\biggr|^2 \nonumber \\
&\quad\qquad{}  + \int \dd x\, \Bigl(\pm \frac{1}{2\kappa m}\rho^2-V(\rho)\Bigr) 
\label{eq:14a}
\end{align}
The $a$ dependence is removed when $\psi(x)$ is replaced by $e^{\frac{i}{2} \int\!\!
\dd \tilde{x}\, \epsilon (x-\tilde{x}) a(\tilde{x})} \psi(x)$, leaving 
\begin{align} 
L &=  \int \dd x\,i \psi^\ast\dot{\psi} -  \frac{1}{2m} \int \dd x\, \bigg|
\Bigl(\partial_x \mp
\frac{1}{2\kappa} \int \dd \tilde{x}\,
\epsilon(x-\tilde{x})\rho(\tilde{x})\Bigr)\psi(x)\bigg|^2
\nonumber \\ 
&\quad{} + \int dx \Bigl(\pm \frac{1}{2\kappa m}\rho^2-V(\rho)\Bigr) 
\label{eq:14b}
\end{align}\label{eq:14}
\end{subequations}%

Finally we choose $V(\rho)$ to be $\pm \frac{1}{2\kappa m}\rho^2$ [this is the same
choice that in (2+1)-dimensions leads to static first-order Bogomolny
equations] and our reduced C-S, $B$-$F$ theory is governed by the
Hamiltonian
\begin{equation}
H = \frac{1}{2m}\int dx \bigg| \Bigl(\partial_x \mp \frac{1}{2\kappa} \int
 d\tilde{x} \epsilon(x-\tilde{x})\rho(\tilde{x})\Bigr)\psi(x)\bigg|^2
\label{eq:15}
\end{equation}
which implies the first-order Bogomolny equation
\begin{subequations}
\begin{equation}
\psi'(x)\mp \frac{1}{2\kappa}\int d\tilde{x}
\epsilon(x-\tilde{x})\rho(\tilde{x})\psi(x)=0
\label{eq:16a}
\end{equation}
solved by 
\begin{equation}\psi(x) = {\rm phase~} \times \frac{\alpha\sqrt{|\kappa|}}{\cosh \alpha
x}
\label{eq:16b}
\end{equation}
\end{subequations}
where $\mp \kappa$ is taken as positive, and $\alpha$ is an integration constant.

On the other hand, we can recognize the dynamics described in (\ref{eq:15}) by
expanding the product:
\begin{align}
 H  = \frac{1}{2m} &\int \dd x\, \Bigl\{ |\psi'|^2 \pm \frac{1}{\kappa}
\rho^2 \Bigr\}
\nonumber \\
+ \frac{1}{24m\kappa^2} &\int \dd x\,  \dd \tilde{x}\, 
\dd  \hat{x}\,  \rho(x) \rho(\tilde{x}) \rho(\hat{x}) \Bigl\{
\epsilon(x-\tilde{x}) \epsilon(x-\hat{x}) +
\epsilon(\tilde{x}-\hat{x})\epsilon(\tilde{x}-x) + \epsilon
(\hat{x}-x)\epsilon(\hat{x}-\tilde{x})\Bigr\} \pp
\label{eq:17}
\end{align}
The last term was symmetrized, leading to a sum of step function products,
which in fact equals to 1.  Consequently the last integral
is $\frac{1}{24m\kappa^2}N^3$, where $N=\int dx \rho(x)$, which is conserved in
the dynamics implied by (\ref{eq:17}).  Hence this term can be removed by
redefining
\begin{equation}
\psi \rightarrow e^{-i \frac{N^2t}{8m\kappa^2}} \psi \pp
\label{eq:18}
\end{equation}
What is left is recognized as the Hamiltonian for the nonlinear Schr\"odinger equation, with
equation of motion
\begin{align}
i\dot{\psi} &=  -\frac{1}{2m} \psi '' -\lambda \rho\psi \nonumber \\
\lambda &\equiv  \mp \frac{1}{m\kappa} \pp
\label{eq:19}
\end{align}

The nonlinear Schr\"odinger equation plays a cycle of interrelated roles in
mathematical physics.  Viewed as a nonlinear, partial differential equation for the
function $\psi$, it is completely integrable, possessing a complete spectrum of
multi-soliton solutions, the simplest of these being the single soliton at
rest.  This requires
$\lambda > 0$, which is always achievable in our reduction by adjustment
of
$\kappa$:
\begin{equation}
\psi_s^{\rm rest} (t,x)= \pm e^{i\frac{\alpha^2}{2m}t} \frac{1}{\sqrt{\lambda
m}}\frac{\alpha}{\cosh
\alpha x} \pp
\label{eq:20}
\end{equation}
Here $\alpha$ is an integration constant, and the result is consistent with (\ref{eq:16b}),
once the redefinition (\ref{eq:18}) is taken into account.  Because of Galileo invariance,
the solution may be boosted with velocity
$v$, yielding
\begin{equation}
\psi_s^{\rm moving} (t,x)= \pm e^{imvx} e^{it (\frac{\alpha^2}{2m} -
\frac{mv^2}{2} )} \frac{1}{\sqrt{\lambda m}} \frac{\alpha}{\cosh \alpha
(x-vt)} \pp
\label{eq:21}
\end{equation}
The soliton solutions can be quantized by the well-known methods of soliton
quantization.  On the other hand, the nonrelativistic field theory can be quantized at
fixed $N$, where it describes $N$ nonrelativistic point particles with pair-wise
$\delta$-function interactions.  This quantal problem can also be solved
exactly, and the results agree with those of soliton quantization.  All these
properties are well known, and will not be reviewed here.\cite{ref:6}

The present development demonstrates that this classical/quantal, completely
integrable theory possesses a Bogomolny  formulation, which is obtained by
using two-dimensional $B$-$F$ gauge theory, which in turn descends from
three-dimensional C-S dynamics.\cite{ref:7}

\subsection{\sl Reduction to modified nonlinear Schr\"odinger equation}

While the previous development started with $B$-$F$ gauge theory, which
descended from a C-S model, and arrived at an interesting (first-order,
Bogomolny) formulation for the familiar nonlinear Schr\"odinger equation, we
now further modify the gauge theory and obtain a novel, chiral, nonlinear
Schr\"odinger equation.

Let us observe first that the above dynamics is nontrivial solely
because we have chosen $V$ to be nonvanishing.  Indeed with $V=0$ in
(\ref{eq:11}), (\ref{eq:12}), and (14), the same set of steps
(removing $B$ and $a$ from the theory) results in a free theory for the $\psi$
field.

To avoid triviality at $V=0$, we need to make the $B$ field dynamically active
by endowing it with a kinetic term.  Such a kinetic term could take the
Klein-Gordon form; however we prefer a simpler expression that describes a
``chiral'' Bose field, propagating in only one direction.  A Lagrange density
for such a field has been known for some time.\cite{ref:8}
\begin{equation}
{\cal L}_{\rm chiral} = \pm \dot{B} B' + v B' B' \pp
\label{eq:22}
\end{equation}
Here $v$ is a velocity, and the consequent equation of motion arising from
${\cal L}_{\rm chiral}$ is solved by $B=B(x \mp vt)$ (with suitable boundary
conditions at infinity), describing propagation in one direction, with velocity
$\pm v$.  Note that $\dot{B} B'$ is \emph{not} invariant against a Galileo boost,
which is a symmetry of $B'B'$ and of (\ref{eq:11}), (\ref{eq:12}), (14):
performing a Galileo boost on $\dot{B} B'$ with velocity $\tilde{v}$ gives rise
to
$\tilde{v} B' B'$, effectively boosting the $v$ parameter in ${\cal L}_{\rm
chiral}$ by $\tilde{v}$. Consequently, one may drop the $v B' B'$
contribution, thereby selecting to work in a global ``rest frame.''  Boosting
a solution in this rest frame produces a solution to the theory with a $B' B'$
term.

In view of this discussion, we supplement the previous Lagrange density
(\ref{eq:11}), (\ref{eq:12}), (14) by $\pm \dot{B} B'$, set $V$ to
zero, and thereby replace (\ref{eq:12}) by
\begin{subequations}
\begin{equation}
{\cal L} = - \kappa \dot{B} \Bigl( a \mp \frac{1}{\kappa} B' \Bigr) + i
\psi^\ast \dot{\psi} - A_0 (\kappa B' + \rho) - \frac{1}{2m} | (\partial_x -
ia) \psi |^2 - \frac{1}{2m} B^2 \rho \pp
\label{eq:23a}
\end{equation}
After redefining $a$ as $a \pm \frac{1}{\kappa} B'$, this becomes equivalent to 
\begin{equation}
{\cal L} = \kappa B \dot{a} + i \psi^\ast \dot{\psi} - A_0 (\kappa B' +
\rho) - \frac{1}{2m} | (\partial_x -
ia \mp i \frac{1}{\kappa} B') \psi |^2 - \frac{1}{2m} B^2 \rho \pp
\label{eq:23b}
\end{equation}
\end{subequations}%
Now we proceed as before: solve Gauss' law as in (13), remove $a$ by
a phase-redefinition of $\psi$, drop the last term in (\ref{eq:23b}) by a
further phase redefinition as in (\ref{eq:18}).  We are then left with 
\begin{equation}
{\cal L} = i \psi^\ast \dot{\psi} - \frac{1}{2m} | (\partial_x 
\pm i \frac{1}{\kappa^2} \rho) \psi |^2 \pp
\label{eq:24}
\end{equation}
It has been suggested that this theory may be relevant to modeling quantum
Hall edge states.\cite{ref:9}

The Euler-Lagrange equation that follows from
(\ref{eq:24}) reads
\begin{equation}
i  \dot{\psi} = - \frac{1}{2m} \Bigl(
\partial_x \pm i\frac{1}{\kappa^2} \rho \Bigr)^2 \psi \pm 
\frac{1}{\kappa^2} j \psi
\label{eq:25}
\end{equation}
where the current density $j$
\begin{equation}
j = \frac{1}{m} {\rm Im} \psi^\ast \Bigl( \partial_x \pm i
\frac{1}{\kappa^2} \rho \Bigr) \psi
\label{eq:26}
\end{equation}
is linked to $\rho$ by the continuity equation 
\begin{equation}
\dot{\rho} + \partial_x j = 0 \pp
\label{eq:27}
\end{equation}
Next we redefine the $\psi$ field by
\begin{equation}
\psi (t,x) = e^{\mp \frac{i}{\kappa^2} \int^{\!\!x}\! \dd y\, \rho (t,y)} \Psi (t,x)
\label{eq:28}
\end{equation}
and see that the equations satisfied by $\Psi$ is
\begin{subequations}
\begin{equation}
i \dot{\Psi} (t,x) \pm \frac{1}{\kappa^2} \int^x \dd y\,  \dot{\rho} (t,y) \Psi
(t,x) = - \frac{1}{2m} \Psi'' (t,x) \pm \frac{1}{\kappa^2} j(t,x) \Psi (t,x) \pp
\label{eq:29}
\end{equation}
But the integral may be evaluated with the help of (\ref{eq:27}), so finally
we are left with\cite{ref:10}
\begin{equation}
i \dot{\Psi} = - \frac{1}{2m} \Psi'' \pm \frac{2}{\kappa^2} j \Psi \pp
\label{eq:30}
\end{equation}
\end{subequations}

This is a nonlinear Schr\"odinger equation similar to (\ref{eq:19}) but with
the current density $j= \frac{1}{m} {\rm Im} \Psi^\ast \Psi'$ replacing the
charge density $\rho = \Psi^\ast \Psi$.  The equation is \emph{not} known to be
completely integrable but it does possess an interesting soliton solution, which
is readily found by setting the $x$-dependence of the phase of $\Psi$ to be
$e^{imvx}$.  Then $j= v\rho$, and our new equation (\ref{eq:30}) becomes the
usual nonlinear Schr\"odinger equation (\ref{eq:19})
\begin{equation}
i \dot{\Psi} = - \frac{1}{2m} \Psi'' \pm \frac{2v}{\kappa^2} \rho \Psi
\label{eq:31}
\end{equation}
that is,  the nonlinear coupling strength of (\ref{eq:19}) is 
\begin{equation}
\lambda = \mp \frac{2v}{\kappa^2} \pp
\label{eq:32}
\end{equation}

The $(\mp)$ sign is inherited from the ``chiral'' kinetic term, see
(\ref{eq:22}), (23); once a definite choice is made (say~$+$),
positive $\lambda$, which is required for soliton binding, corresponds to
definite sign for $v$ (say positive); that is,  the soliton solving
(\ref{eq:31}) moves in only one direction.  Explicitly, with the above choice
of signs, the one-soliton solution reads
\begin{equation}
\Psi_s (t,x) = \pm e^{imvx} e^{it \bigl( \frac{\alpha^2}{2m} -
\frac{mv^2}{2} \bigr)} \frac{\kappa}{\sqrt{2mv}} \frac{\alpha}{\cosh
\alpha (x-vt)} \pp
\label{eq:33}
\end{equation}
We see explicitly that $v$ must be positive; the soliton cannot be brought to
rest; Galileo invariance is lost.  

The characteristics of the solution are as follows:
\begin{equation}
N_s =  \frac{\alpha \kappa^2}{mv} \pp
\label{eq:34}
\end{equation}
The energy is obtained by integrating the Hamiltonian.
\begin{equation}
E = \int \dd x\, \frac{1}{2m} |\Psi'|^2 \pp
\label{eq:35}
\end{equation}
and on the solution (\ref{eq:33}), takes the value appropriate to a massive,
nonrelativistic particle.
\begin{equation}
E_s = \half M_s v^2
\label{eq:36}
\end{equation}
where
\begin{equation}
M_s = mN_s \bigl(1 + \frac{1}{3 \kappa^4} N^2_s\bigr) \pp
\label{eq:37}
\end{equation}
The conserved field momentum in this theory reads
\begin{equation}
P = \int \dd x\, \bigl(m j + \frac{1}{\kappa^2} \rho^2 \bigr)
\label{eq:38}
\end{equation}
and on the solution (\ref{eq:33}) its value again corresponds to that of a
massive, nonrelativistic particle
\begin{align}
P_s &= M_s v
\label{eq:38_2}\\
E_s &= \frac{P_s^2}{2M_s}
\label{eq:39}
\end{align}

As already remarked, the model is not Galileo invariant, but one can verify
that it is scale invariant.  Indeed one can show that the above kinematical
relations are a consequence of scale invariance.\cite{ref:10}

The soliton solution (\ref{eq:33}) can be quantized; also the quantal many-body
problem, which is implied by (\ref{eq:24}), can be analyzed.  Because the system does
not appear integrable, exact results are unavailable, but one verifies that at weak
coupling, the two methods of quantization (soliton, many-body) produce identical
results.\cite{ref:10,ref:11}

\section{Reducing the Chern-Simons Term Using a Symmetry}

Reducing a three-dimensional C-S term in (\ref{eq:1jp}) by a symmetry yields another
topologically interesting structure.\cite{jp:12} Specifically, reducing by radial
symmetry results in a  one-dimensional quantum mechanical model, which has recently
been used in an analysis of finite-temperature C-S theory.\cite{jp:13}  Earlier
calculations seem to indicate that the coefficient of the induced C-S term depends on
the temperature,\cite{jp:14} contradicting that it must have quantized,
discrete values.  The puzzle became resolved once it was realized that finite temperature
calculations to fixed perturbative order necessarily violate gauge invariance, which is
restored only after all orders are summed.\cite{jp:13,jp:14b}  (At zero
temperature, finite-order calculations suffice to exhibit the complete, induced C-S
term.\cite{jp:15})

The all-order summation was first accomplished in a toy quantum mechanical model,
which had been introduced a decade earlier for the purpose of exhibiting in a simple
setting some of the peculiar topological/geometrical effects of quantized C-S
theory.\cite{jp:16}  We shall show that this model is not merely a pedagogical toy; in
fact, it coincides with the three-dimensional C-S term, reduced by radial symmetry.

Consider the C-S action in three-dimensional space.
\begin{equation}
NW(A) = -\frac{N}{4\pi} \int \dd^3 x \epsilon^{ijk} \Tr 
(\partial_i A_j A_k  + \two3 A_i A_j A_k)
\label{eq:40jp}
\end{equation}
The coefficient of $W(A)$ is chosen so that the quantization condition is obeyed with
integer $N$.  Let us consider SU(2) case and take for $A_i^a$ the radially symmetric
\emph{Ansatz}, familiar form, from monopole/instanton studies.
\begin{equation}
A_i^a = (\delta_i^a - \hat{r}^i\hat{r}^a) \frac{1}{r} \psi_1 + \epsilon^{iaj} \hat{r}^j 
\frac{1}{r} (\psi_2-1) + \hat{r}^i \hat{r}^a A
\label{eq:41jp}
\end{equation}
Here $\psi_m, m=1,2$ and $A$ are functions just of $r$.  Substituting (\ref{eq:41jp})
into (\ref{eq:40jp}), performing the angular integral, leaves
\begin{eqnarray}
NW(A) &=& N\int_0^\infty \dd r \big(\epsilon^{mn} \psi_m (D\psi)_n - A \big) 
\label{eq:42jp} \\
(D\psi)_m &\equiv& \psi_m' - \epsilon_{mn} A\psi_n
\label{eq:43jp} 
\end{eqnarray}
Here the dash denotes $r$-differentiation and we have dropped an end point
contribution, $\psi_1 \big|_{r=0}^{r=\infty}$.  Note that (\ref{eq:42jp}) is a
(0+1)-dimensional field theory, that is,  a quantum mechanical system, except that
$r$ has only half the range of $t$.  The system has local $U(1)$ gauge symmetry, with a
one-dimensional C-S term: $A$.  Under a gauge transformation, the action changes by
$\Delta I = -N \Delta \theta$, where $\Delta \theta = \int \dd \theta$; when $\Delta
\theta$ is restricted to an integral multiple of 2$\pi$, gauge invariance of $e^{iI}$ is
assured by the integer $N$.\cite{jp:16}

The recent analysis of the C-S term at finite temperature~\cite{jp:13, jp:15} was
based on the model (\ref{eq:42jp}).

Finally, we wish to mention that the radially symmetric C-S term (\ref{eq:42jp}) may
be inserted into the classical field equation of the Yang-Mills/Higgs model in
(3+1)~dimensions, for which static 't\thinspace Hooft-Polyakov monopole solitons exist,
with explicitly known profiles in the Bogomolny-Prasad-Sommerfield limit.  These
static solutions can also be viewed as instantons of the same theory in
(2+1)~dimensions, continued to Euclidean 3-space.  The C-S term in the (3+1) theory
violates Lorentz invariance in an interesting, gauge invariant fashion.\cite{jp:17}  On the
other hand, when the C-S term is added to the Euclidean three-dimensional theory, it enters
with an imaginary coefficient, which is inherited from the continuation to imaginary
time of the
(2+1)-dimensional model.  Nevertheless, the final equations are real.  One can show
that the addition of the topological C-S interaction destroys the topological
excitations; the equations no longer admit monopole or instanton solutions.~\cite{jp:12}

\end{document}